# Optimization of the high-frequency magnetoimpedance response in melt-extracted Co-rich microwires through novel multiple-step Joule heating

O. Thiabgoh[1,2]*, T. Eggers[1], C. Albrecht[1], V.O. Jimenez[1], H. Shen[3], S.D. Jiang[3], J. F. Sun[3], D.S. Lam[4], V.D. Lam[4], and M.H. Phan[4,]*

[1] Department of Physics, University of South, Florida, Tampa, Florida 33620, USA

[2] Department of Physics, Faculty of Science, Ubon Ratchathani University, Warin Chamrap, Ubon Ratchathani 34190, Thailand

[3] School of Materials Science and Engineering, Harbin Institute of Technology, Harbin 150001, P. R. China

[4] Institute of Materials Science, Vietnam Academy of Science and Technology, 18 Hoang Quoc Viet, Ha Noi, Vietnam

The optimization of high frequency giant magnetoimpedance (GMI) effect and its magnetic field sensitivity in melt-extracted $Co_{69.25}Fe_{4.25}Si_{13}B_{12.5}Nb_1$ amorphous microwires, through a multi-step Joule annealing (MSA) technique, was systematically studied. The surface morphology, microstructure, surface magnetic property, and high frequency GMI response of the Co-rich microwires were explored using scanning electron microscopy (SEM), magneto-optical Kerr effect (MOKE) magnetometry, transmission electron microscopy (TEM), and impedance analyzer, respectively. An initial dc current ($i_{dc}$) of 20 mA, which was then increased by 20 mA at every time-step (10 min) up to 300 mA, was applied to the microwires. The MSA of 20 mA to 100 mA remarkably improved the GMI ratio and its field sensitivity up to 760% (1.75 time of that of the as-prepared), and 925%/Oe (more than 17.92 times of that of the as-prepared) at an operating frequency of 20 MHz, respectively. Our study indicates that the MSA technique can enhance the microstructures and the surface magnetic domain structures of the Co-rich magnetic microwires, giving rise to the GMI enhancement. This technique is suitable for improving the GMI sensitivity at small magnetic fields, which is highly promising for biomedical sensing and healthcare monitoring.

Corresponding authors: thiabgoh.ongard@gmail.com (O.T.); phanm@usf.edu (M.H.P.)

**1. Introduction**

The development of Co-rich microwires with magnetic softness and giant magnetoimpedance (GMI) effect has enabled a new class of magnetic sensors with extremely high field sensitivity down to the pico-Tesla regime at room temperature [1-2]. Thanks to their superior properties, i.e., excellent mechanical properties, broad frequency response, thermal stability, and very high sensitivity, the GMI sensors based on Co-rich microwires have recently been used for detection of biomagnetic fields [3], magneto-LC resonance sensing [4], and detection of magnetically labeled bioanalytes [5]. Therefore, this sensor technology is highly promising for small magnetic field detection, biosensing and real-time healthcare monitoring [1,2,4,5]. In order to achieve such ultra-high sensitivity, fast speed response, miniaturized size, and stability, a number of recent studies have focused on improving the GMI effect and its magnetic field sensitivity in these microwires through thermal treatment and chemical doping [6-9].

Joule annealing Co-rich microwires as a post-processing treatment has revealed the improvement of GMI properties [10,11]. In a simple Joule annealing scheme, the sudden application of dc current intensity of ~ 100 mA for an appropriate time showed an overall improvement of the GMI response [12]. This is due partially to the relief of quenched-in stresses, structural relaxation, and microstructure evolution during Joule heating [13]. In addition, the circumferential magnetic domain structure was observed to be enhanced by the self-induced magnetic field of the annealing current [14,15]. Furthermore, the formation of nanocrystals in the amorphous matrix caused by the heat treatment increases the electrical conductivity [12]. The Joule heating technique has more advantages than conventional annealing methods and promotes the high frequency GMI effect in Co-rich microwires. However, an asymmetric peak in the GMI response, attributed to induced helical domains, was observed when the current intensity is greater than 100 mA. To combat this, a stepped approach to the highest current intensity has shown a GMI enhancement up to 360% and reduced the undesired asymmetric GMI peak [16]. Furthermore, this multi-step Joule-heating markedly promote the field sensitivity for the weak field regime, $H < 2$ Oe. However, most results were explored for the excitation frequency below 20 MHz [12,15,16]. Despite recent investigations have shown great promotion in the GMI-ratio, a clear evident between GMI-promotion through multi-step current annealing and their surface properties for high frequency GMI-response is lacking.

In the present study, the high frequency GMI response in melt-extracted amorphous $Co_{69.25}Fe_{4.25}Si_{13}B_{12.5}Nb_1$ microwires over a high frequency range (1 MHz - 1 GHz) has been studied. A comparative study of single- and multi-step Joule-heating in the Co-rich microwires was performed. We found that tailoring the GMI effect through the multi-step annealing technique enhanced GMI-ratio and field sensitivity to 760% and 925%/Oe, respectively. We demonstrate that the multi-step current annealing promotes the low magnetic field sensitivity. The results are potentially applicable to the development of highly sensitive, high operating frequency GMI-based sensors.

**2. Experimental**

The high quality amorphous magnetic microwires with a nominal composition of $Co_{69.25}Fe_{4.25}Si_{13}B_{12.5}Nb_1$ (in wt%) were fabricated by a melt-extraction technique as previously described [17]. The microstructure of the as-prepared and annealed microwires was characterized by high-resolution transmission electron microscopy (HRTEM) (Tecnai G2F30). The surface morphology and the nominal elemental composition of the samples were investigated by scanning electron microscopy (SEM) and energy dispersive spectroscopy (EDS) (JEOL JSM-6390LV), respectively. Surface magnetic properties of the samples were characterized using Magneto-optical Kerr effect (MOKE) magnetometry as previously described in our study [18]. The as-prepared microwires with diameter ~55 μm and 10 mm in length ($l$) were selected from a strand of $Co_{69.25}Fe_{4.25}Si_{13}B_{12.5}Nb_1$ microwires. Identical samples ($d$~55 μm, $l$ ~10 mm), as determined by an initial magneto-impedance measurement, were chosen for the comparative post-heating treatments. The selected microwire was soldered to SMA ports with copper ground plane. The mounted microwires were subjected to a constant dc current of $i_{dc}$ = 100 mA for $t$ = 10 min. This annealing scheme is denoted as single-step annealing (SSA). For comparison, a multi-step annealing scheme (MSA) is applied to the second microwire consisting of an initial current intensity $i_{dc}$ = 20 mA and then increasing the current intensity by $i_{step}$ = 20 mA every $t_{step}$ = 10 min until $i_{dc}$ = 300 mA. A schematic for sample treatments using SSA and MSA is shown in Fig. 1. The GMI response over the high frequency range ($f_{ac}$ = 1 MHz-1 GHz) was measured using a HP4191A impedance analyzer through a transmission line [19]. Standard calibration procedures (short, open, and 50 Ω) were performed before the measurement. The GMI response over the high frequency range ($f_{ac}$ = 1 MHz-1 GHz) was measured using a HP4191A impedance analyzer through a transmission line [19]. Standard calibration procedures (short, open, and 50 Ω) were performed before the measurement. All measurements were performed at room temperature. The HP4191A determines the complex reflection coefficient (Γ) of a measurement frequency test signal applied to the terminated transmission line with a reference impedance $z_0 = 50\ \Omega$ that matches the input impedance of the analyzer. The complex impedance of a test sample is determined by

$$Z = z_0 \left(\frac{1+\Gamma}{1-\Gamma}\right) = R + jX, \tag{1}$$

where $R$ is the resistance, $X$ is the reactance, and $j$ is the imaginary unit. For each frequency measurement, a pair of Helmholtz coils was employed to generate the external field ($H_{dc}$) along the longitudinal direction of the microwires. The GMI ratio (ΔZ/Z) is defined as

$$\frac{\Delta Z}{Z}(\%) = \frac{Z(H) - Z(H_{\max})}{Z(H_{\max})} \times 100, \tag{2}$$

where $Z(H)$ is the impedance at the field $H$, and the $H_{\max}$ represents the maximum induced magnetic field by the Helmholtz coil ($H_{\max}$ = 115 Oe), respectively. In the present study, the magnetic field sensitivity of GMI ($\eta$) is defined as

$$\eta = \frac{d}{dH}\left(\frac{\Delta Z}{Z}\right). \tag{3}$$

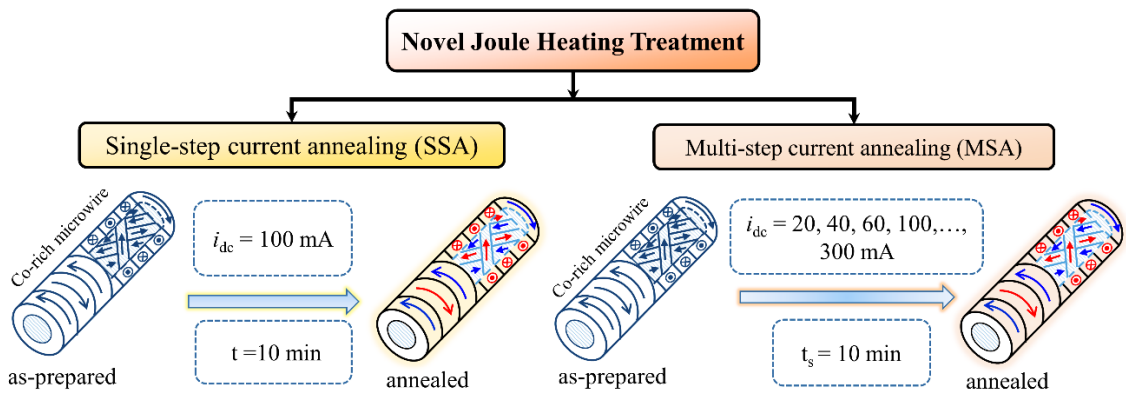

**Fig. 1** Schematic for sample treatments using SSA and MSA methods.

## 3. Results and discussion

### *3.1 Structural characterization*

The surface microstructure of a soft magnetic microwire plays an essential role in high frequency GMI properties due to the large skin effect [9]. To explore the surface morphology and microstructure, SEM micrograph, EDS spectra, XRD pattern, and HRTEM image of as-prepared $Co_{69.25}Fe_{4.25}Si_{13}B_{13.5}Nb_1$ microwires are shown in Fig. 2(a)-(d), respectively. As illustrated, the high quality microwire has a smooth surface and cylindrical shape over most of the wire. The composed elements of the microwire are presented in the EDS spectra. The EDS spectra show 69.3 wt% of Co (used as a normalized element), 4.6 wt% of Fe (Sd. = 0.10), 14.4 wt% of Si (Sd. = 1.1), and 1.0 wt% of Nb (Sd. = 1.10). The XRD pattern (Fig. 2(c)) has a broad peak feature which indicates α-CoFe-phase embedded in an amorphous matrix with center near $2\theta = 45$ deg. The high-resolution TEM images exhibits the presence of nanocrystal-like phase embedded in the amorphous matrix. The corresponding electron diffraction pattern (Fig. 2(d) inset) exhibits a diffused broad ring indicating random distribution of short-range order in the as-prepared microwires.

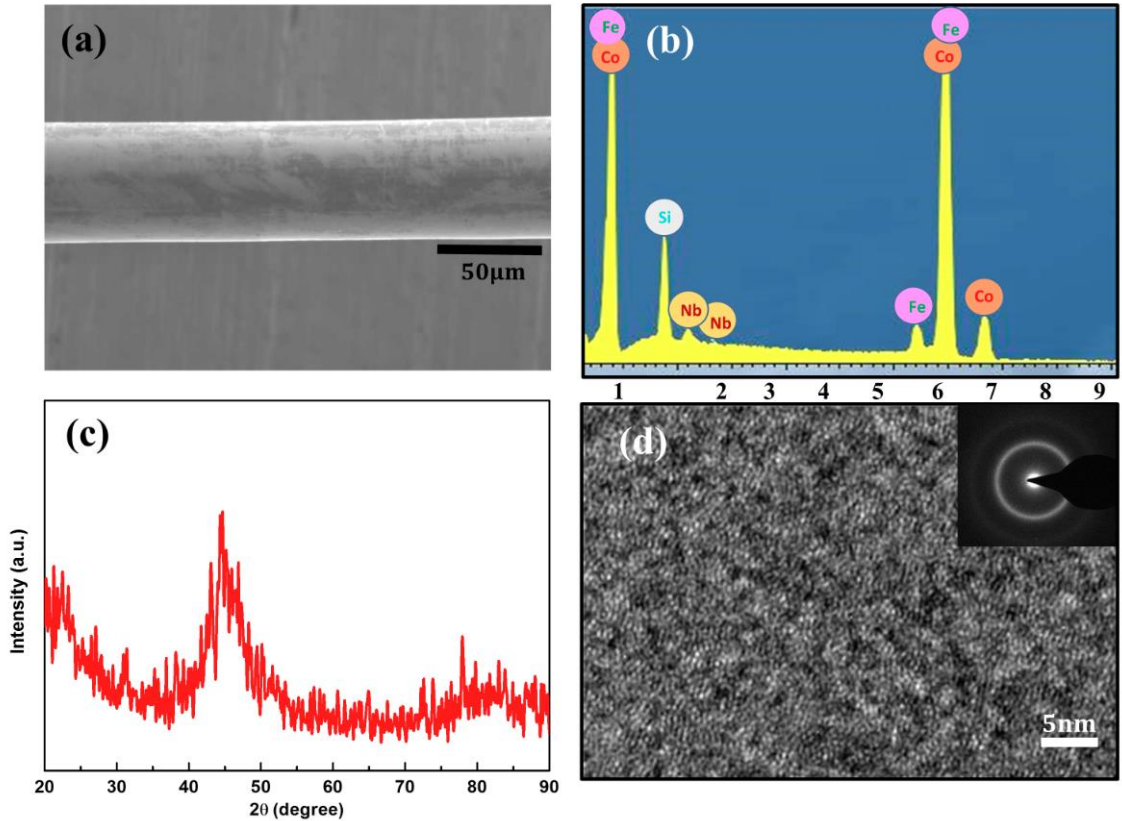

**Fig. 2** (a) The SEM micrograph, (b) EDS spectra, (c) XRD pattern, and (d) HRTEM image of as-prepared $Co_{69.25}Fe_{4.25}Si_{13}B_{13.5}Nb_1$ microwires. Insets in (d) shows the corresponding electron diffraction pattern corresponding to the HRTEM.

The surface morphology of the as-prepared wires at different stages of the annealing schemes is shown in Fig. 3(a)-(d). A smooth surface is observed in the as-prepared, SSA@100mA, and MSA@100mA annealed samples. However, crystalline pieces (>100 nm) on the surface are clearly observed after the MSA@200mA (Fig. 3(d)). This is due to the temperature experienced during current intensity exceeding the crystallization temperature of the Co-rich microwire, which is typically ~ 753 K for the Co-rich microwires [20,21].

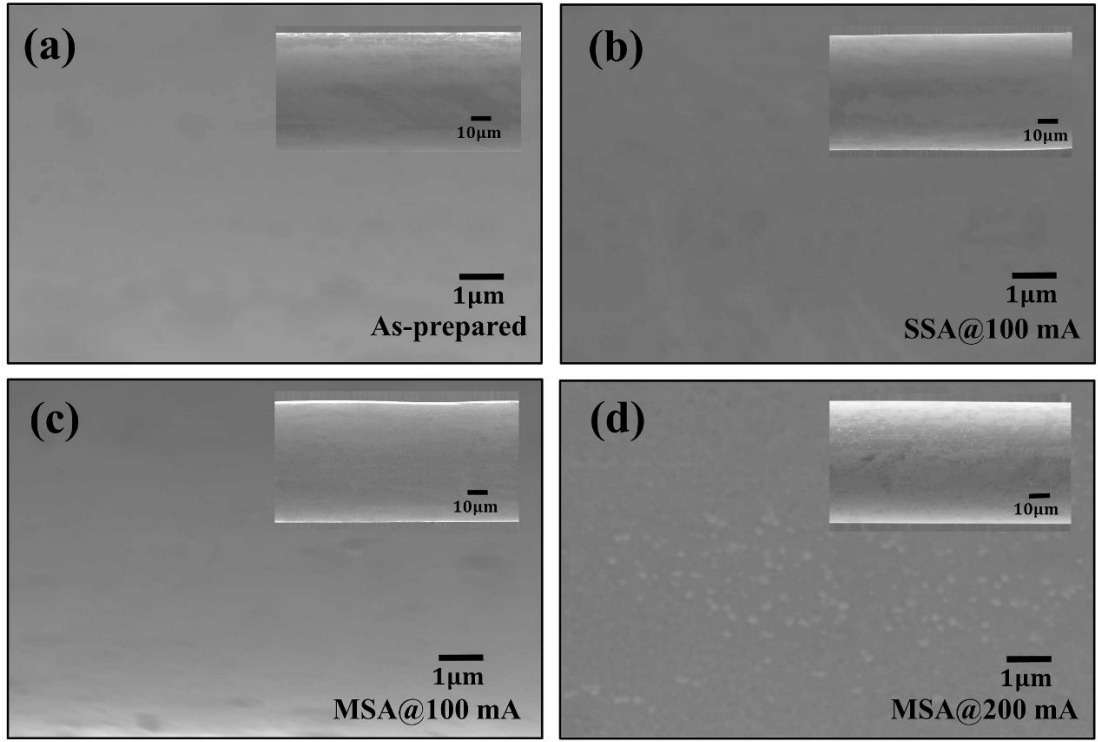

**Fig. 3** (a) The SEM micrograph of the as-prepared the Joule annealing microwires; (b) SSA@100mA, (c) MSA@100mA, and (d) MSA@200mA.

It is well known that applying small current ($i_{dc}$ < 100 mA) through amorphous microwires does not change the composition of their composed elements [12,16]. In contrast, subjecting a high current ($i_{dc}$ ~ 100-200 mA), thereby heating the microwire to above the glass transition temperature ($T_g$), could introduce crystallization into the microwire [22]. Table 1 shows the EDS measurement from the as-prepared, SSA@100mA, and MSA@100mA microwire. As can be seen, the standard deviation of elemental composition is less than 1%. This conforms that the heat treatment (SSA@100mA, and MSA@100mA) does not change the nominal composition at the sample surface. It is worth mentioning that the composed elements become significantly alter when the application of dc annealed current is high up to 200 mA.

**Table 1** the EDS qualitative data for the nominal elements in the Co-rich magnetic microwires.

| Sample | Composition for the nominal elements | | | | |
|---|---|---|---|---|---|
| | $Si_{13}$ | $Fe_{4.25}$ | $Co_{69.25}$ | $Nb_1$ | $B_{12.5}$ |
| As-cast | 14.42 | 4.41 | 69.25 | 1.21 | 10.70 |
| SSA | 15.05 | 4.47 | 69.25 | 1.31 | 9.92 |
| MSA | 15.19 | 4.49 | 69.25 | 1.32 | 9.75 |
| Ave. | 14.89 | 4.46 | 69.25 | 1.28 | 10.12 |
| Sd. | 0.41 | 0.04 | 0.00 | 0.06 | 0.51 |

The microstructure of the as-prepared (Fig. 4(b)) and Joule heated microwires, investigated using HRTEM is presented in Fig. 4. The HRTEM image of the as-prepared wire, Fig. 4(b), is typical for this material, which is characterized as containing randomly oriented "nanocrystallites" or crystal nuceli within an amorphous matrix. No long-range ordering of the nanocrystallites is observed in the as-prepared wire, which is confirmed by the main blurry halo of the SAED pattern (Fig. 4(b), inset). Similar results have been observed in Co-rich microwires with different compositions as well [12,16,23]. After the SSA@100mA annealing protocol, the TEM image (Fig. 4(c)) shows a similar random distribution of nanocrystallites however the SAED shows some additional diffuse rings, indicating nucleation of crystallites at the surface due to the SSA@100mA procedure. However, after the MSA@100mA treatment, the electron diffraction pattern of Fig. 4(d) inset produced crystalline diffraction bright spots on top of the diffused broad ring pattern and in addition some outer rings in Fig. 4(c) inset do not appear. Therefore, the MSA treatment produces a nanocrystalline-embedded matrix with greater degree of uniformity in phase and orientation of the formed nanocrystals [12,16].

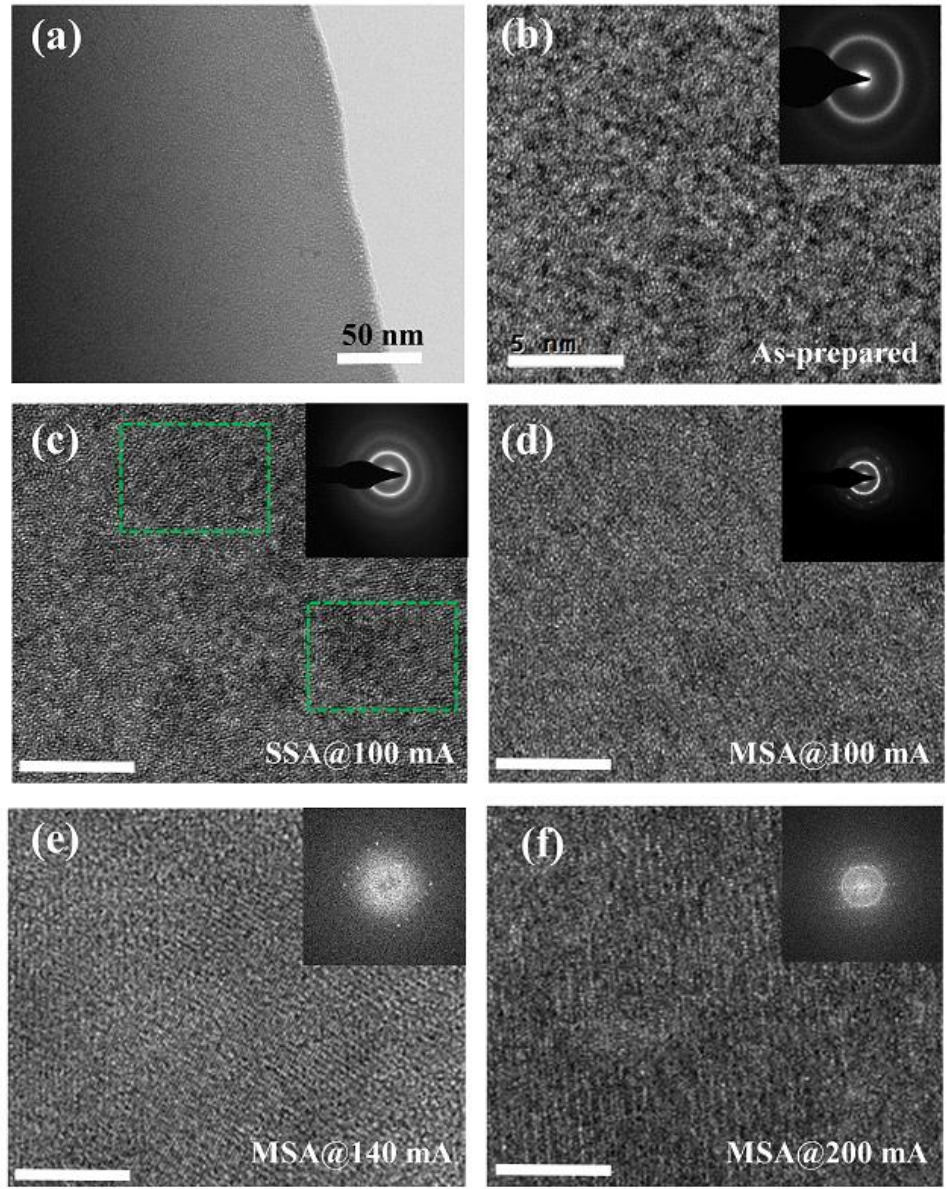

**Fig. 4** The HRTEM images of (a) a prepared sample, (b) as-prepared, (c) SSA@100mA, (d) MSA@100mA (e) MSA@140mA and (f) MSA@200mA microwires. Insets show the corresponding electron diffraction patterns (b-d) and selected area electron diffraction (SAED) (e-f) of the investigated samples.

The formation of significant nanocrystalline size is presented when the applied current intensity reaches $i_{dc}$ = 140 mA. As can be seen in Fig. 4(e), the TEM image shows large nanocrystalline size (> 5 nm) in the amorphous matrix. The electron diffraction displays high ordering of poly-nanocrystalline phases in the annealed sample (MSA@140mA). Thus, this annealing parameter can introduce large nanocrystalline size and increase the volume fraction of the nanocrystalline phases into the magnetic microwires. More importantly, the application of the current up to 140 mA has reached the crystallization temperature of the Co-rich microwire [17]. In addition, the clear crystalline phase is observed in the MSA@200mA (T~1000 K), which confirms the result obtained from the SEM. This introduced large crystalline phase in amorphous magnetic materials could have great impact on their magnetic and GMI properties [14].

*3.2 Surface magnetic properties.*

Hysteresis loops obtained from MOKE is a powerful tool to characterize the surface magnetic properties of materials. To understand the surface magnetic properties of the Co-rich microwire, longitudinal MOKE was performed. In this measurement, the magnetic field ($H_{ex}$) was applied in axial direction; therefore, the M-H loop behavior represents the magnetization along the length of the wire [24-26]. Figure 5(a) demonstrates the MOKE hysteresis loops obtained from the as-prepared microwire. The surface hysteresis loop shows rectangular-like shape with coercivity of $H_c$ = 0.25. The small $H_c$ suggests that the rapid change in the surface magnetization occurred by domain wall movement. A similar experimental result has been reported by Chizhik *et al.* [26]. The square M-H loop indicates the surface domain structures of the microwires are irregular oriented along the axial direction of the microwires [26]. As a result of the SSA@100mA protocol (Fig. 5(a)), the coercivity increased to $H_c$ = 0.75 Oe of in the SSA@100mA microwire and it became harder to magnetize in the axial direction. The increase of coercivity may come from the induced magnetic nanocrystalline phases into the amorphous matrix. It is worth mentioning that the effective anisotropy field ($H_k$) estimated by the saturation point in the hysteresis loop increases about two times that of the as-prepared microwires.

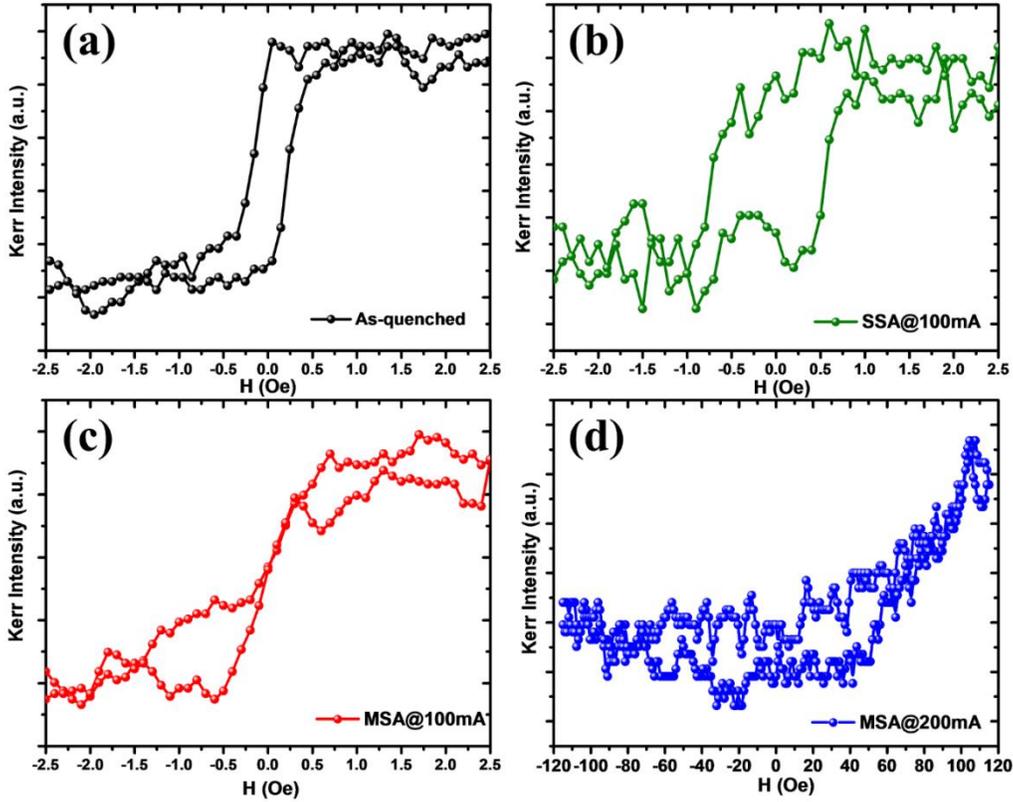

**Fig. 5** The measured Hysteresis loops using longitudinal MOKE for (a) the as-prepared, (b) SSA@100mA, (c) MSA@100mA, and (d) MSA@200mA microwires.

Furthermore, recent studies have shown that the application of dc currents to a magnetic microwire promotes the surface magnetic properties of the microwires [7,12,16]. As seen from Fig. 5(c), the dramatic change of the MOKE hysteresis loop was observed from the MSA@100mA sample. This hard axis loop of the surface magnetization suggests that the SMDs are mostly oriented in the circumferential direction [7,24,25]. The hard-axis surface magnetization process is associated with the rotation of surface magnetic domains into the axial field direction [24,25,27]. In this case, the coercivity is absent but larger $H_k$ is observed, which is due to the strong circumferential anisotropy induced by the MSA@100mA treatment. It is worth mentioning that, the application of high dc current exceeded $T_x$ to the Co-rich microwires will induce crystalline phase and deteriorate the circumferential magnetic domain structures. As a result, the magnetic hardening would be observed as shown in Fig. 5(d).

*3.3 High-frequency GMI effect*

It is established that the GMI-ratio and its field sensitivity of magnetic microwires strongly depends on their excitation frequencies [11]. To compare the impacts of suddenly (SSA) and gradually (MSA) heating to the microwires, the enhanced GMI ratio and $\eta$ between the SSA and MSA are shown in Fig. 6. As can be seen, the MSA@100mA greatly promotes the GMI ratio and $\eta$ over the high frequency range as well as overcome the SSA@100mA does. In our previous studies, the single-step Joule annealing (SSA) by applying current 5 mA for 20 minutes promotes the GMI ratio and $\eta$ in Co-rich microwires to 610% and 500%/Oe (at $f_{ac}$~40 MHz ), respectively [7]. In this experiment the maximum GMI ratios were found at $f_{ac}$ = 20 MHz, which are 500% (as-prepared), 650% (SSA@100mA) and 760% (MSA@100mA). The maximum $\eta$ for the as-prepared microwire is 179.55 %/Oe for $f_{ac}$ ~80 MHz. The SSA@100mA and MSA@100mA samples reach their maximum $\eta$ at $f_{ac}$ ~20 MHz, which are 619 %/Oe and 925%/Oe, respectively. As mentioned earlier, the application of dc currents (80-100 mA) for a suitable time promotes the GMI-ratio and the filed sensitivity in the Co-rich microwires [12,16].

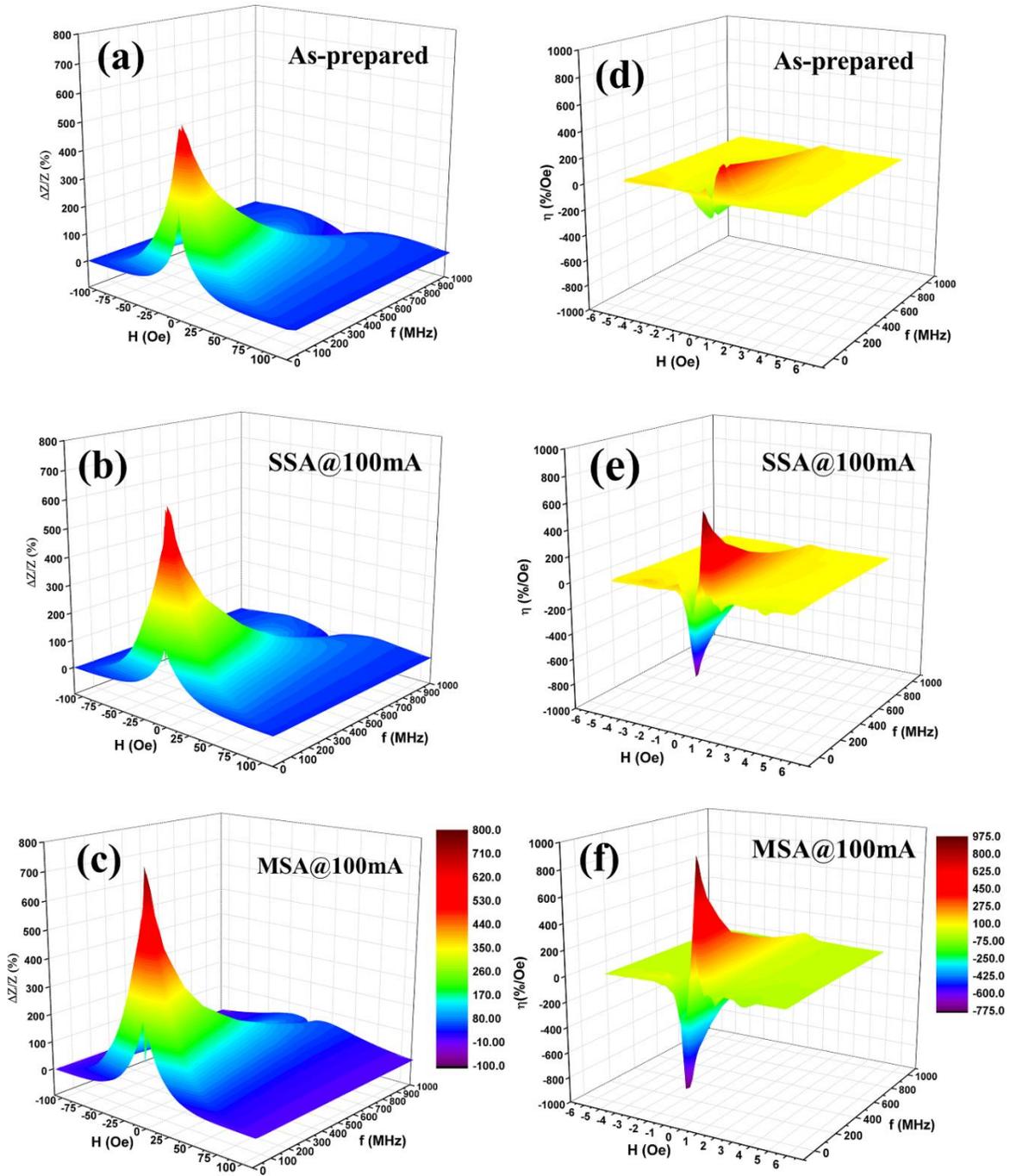

**Fig. 6** The field dependence of MI response and its field sensitivity over wide frequency (1 MHz- 1 GHz). (a), (d) as-prepared, (b), (e) SSA@100mA, (c),(f) MSA@100mA, respectively.

In appreciation of the low field enhancement of the GMI ratio, the sharpness of the dip in the GMI-response at near-zero field at 20 MHz for the as-prepared and annealed samples

(SSA@100mA, MSA@100mA) are shown in Fig. 7(a)-(b). As can be seen, the MSA@100mA shows significant improvement in the GMI ratio at 20 MHz. The maximum GMI ratio is 760%, which is 1.75 times of 500% (as-prepared) and 1.17 times of 650% (SSA@100mA), respectively. More importantly, there is a remarkable improvement of $\eta$ for the MSA@100mA to 925%/Oe, which is 17.92 times of 51.61 %/Oe (as-prepared) and 1.49 times of 619 %/Oe (SSA@100mA), respectively. A similar approach has been employed to improve the GMI-ratio in $Co_{68.15}Fe_{4.35}Si_{12.25}B_{13.75}Nb_1Cu_{0.5}$ and $Co_{68.15}Fe_{4.35}Si_{12.25}B_{12.75}Zr_3$ microwires [12,16]. However, the studies focused on $f_{ac}$ up to 20 MHz and the reported GMI ratio and $\eta$ are 364% ($\eta$ ~ 585.71 %/Oe) and 256% ($\eta$ ~ 372.57 %/Oe), respectively. Therefore, this multi-step current annealing method is more effective in promoting the high frequency GMI effect in Co-rich microwires than the single step current annealing technique.

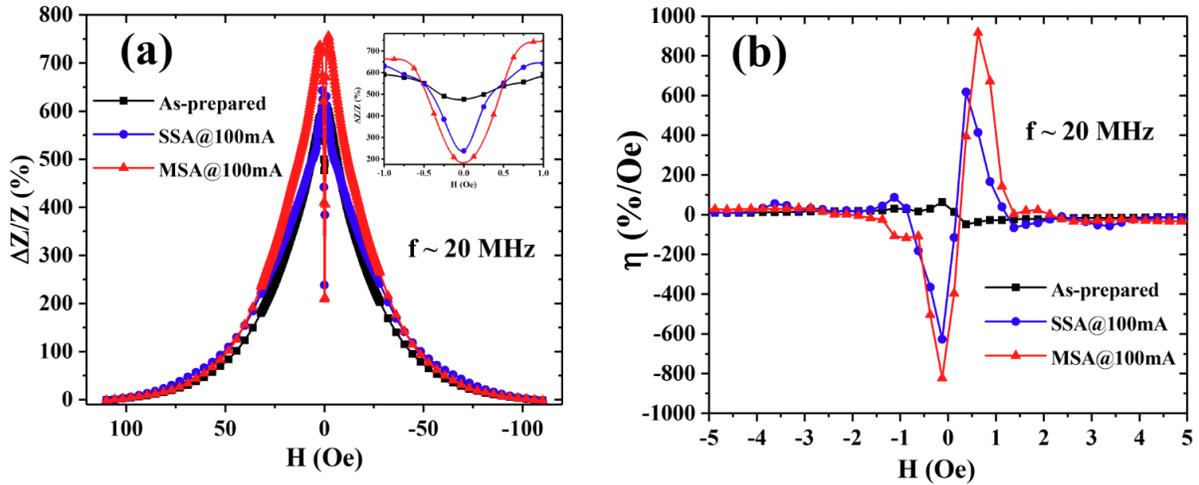

**Fig. 7** (a) The field dependence of the MI and (b) field sensitivity of as-prepared (black) and annealed samples; SSA@100mA (blue), MSA@100mA (red) at 20 MHz.

In order to correlate the GMI effect and the microstructure evolution owing to the multi-step Joule annealing, the maximum GMI ratio, $[\Delta Z/Z]_{max}$ and field sensitivity ($\eta$) of the annealed microwires as a function of annealing current ($i_{dc}$) are shown in Fig. 8(a). As can be seen, the GMI-ratio gradually increases with increasing current $i_{dc}$ up to 80 mA. Then, the ratio drastically changes to the highest value (760%) at $i_{dc}$ ~100mA. A similar development of the $\eta$ is also observed. As mentioned previously, the annealing current below 80 mA affects the structural relaxation and reduces the structural stress from the as-prepared state. This structural relief may contribute to the increases of the effective permeability and electrical conductivity of the microwires [12]. Thus, the GMI-ratio is promoted. Furthermore, once the annealing current is close to 100mA, the nanocrystalline becomes significant and rational to enhance the electrical and soft magnetic properties of the microwires, which is mentioned previously [16,23]. The circumferential self-induced magnetic field promotes well define SMDs [15,16]. Nonetheless, the thermal energy due to Joule heating promotes nano-crystalline in the amorphous matrix. Therefore, the magnetic permeability of the microwires becomes larger (softer) and the electrical conductivity becomes smaller in comparing to the as-quenched microwires. Therefore, the GMI-ratio and $\eta$ simultaneously reach the maximum value at this point.

The decrease in GMI-ratio and $\eta$ is observed when annealing current is higher than 100 mA. This is due to the nanocystalline size and its volume fraction become large and deteriorate soft magnetic properties of the microwire [12,23]. In particular, the annealing current up to 140 mA introduced poly-nanocrystalline phase in to the microwire structure shown previously in the Fig. 4(e). Furthermore, the large crystalline sizes are formed when the annealed current generating heat above the $T_x$ ($i_{dc}$ ~200 mA) shown in Fig. 4(f). The effective magnetic permeability of the microwires decreases due to the introduction of large magnetic crystalline anisotropy into the microwire [12,14]. Therefore, the GMI-ratio and $\eta$ markedly deteriorate.

The effective magnetic anisotropy field ($H_{k,eff}$) estimated by GMI of the microwires as a function of annealing current is shown in Fig. 8(a). The $H_{k,eff}$ does not change until $i_{dc} = 140$ mA.

The jump in $H_k$ corresponds to the large volume fraction of nanocrystalline phase is introduced into the microwire as seen in the TEM image and electron diffraction. The $H_k$ further increases up to 6 Oe when $i_{dc}$ = 200 mA ($T \sim 1000$ K). At this current intensity, the microwire's temperature exceeds the glass transition temperature of a Co-rich microwire and it becomes primarily crystalline and the deterioration of GMI properties is clearly observed.

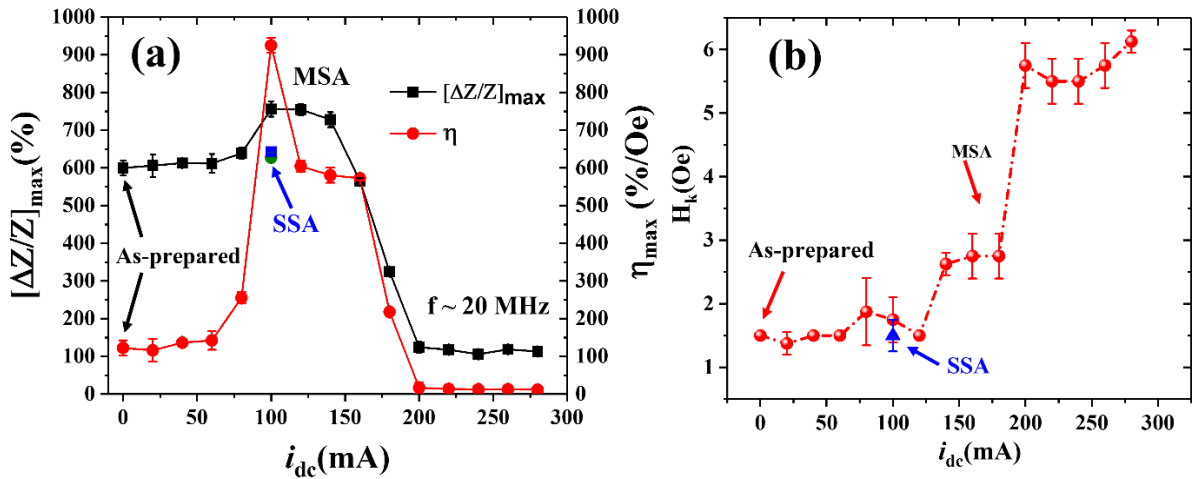

**Fig. 8** (a) Annealed currents dependence of GMI-ratio (black square) and field sensitivity of as-cast and annealed samples; SSA@100 mA (green and blue squares), MSA@20-280 mA (red circle). (b) Annealed currents and anisotropy field ($H_k$) of the annealed samples at 20 MHz. (b) SSA@100mA, (c) MSA@100mA and (d) MSA@200mA.

## 4. Conclusion

The multi-step Joule annealing technique has been employed to optimize the high frequency GMI effect and its magnetic field sensitivity in melt-extracted amorphous $Co_{69.25}Fe_{4.25}Si_{13}B_{12.5}Nb_1$ microwires. We found that increasing the $i_{dc}$ from 20 mA to 100 mA for 10 minutes remarkably improved the GMI ratio and its field sensitivity up to 760%, and 925%/Oe at $f \approx 20$ MHz, respectively. The multi-step Joule annealing method is more advantageous than the single-step

Joule annealing one on account of the enhancement of the microstructures and SMDs of the Co-rich microwires. This alternative annealing technique is appropriate for improving the magnetic field GMI sensitivity at small magnetic fields, which is very promising for biomedical sensing and healthcare applications.

**Acknowledgments**

Research at the University of South Florida was supported by the U.S. Department of Energy, Office of Basic Energy Sciences, Division of Materials Sciences and Engineering under Award No. DE-FG02-07ER46438. This research was also supported by Vietnam Ministry of Science and Technology through the national-level project ĐTĐLCN.17/19.

**References**

[1] M. H. Phan, H. X. Peng, Giant magnetoimpedance materials: Fundamentals and applications, Prog. Mater. Sci., 53 (2008) 323-420.

[2] T. Uchiyama, K. Mohri, Y. Honkura, and L. V. Panina, Recent advances of pico-tesla resolution magneto-impedance sensor based on amorphous wire CMOS IC MI sensor, IEEE Trans. Magn. 48 (2012) 3833-3839.

[3] S. Nakayama, T. Uchiyama, Real-time measurement of biomagnetic vector fields in functional syncytium using amorphous metal. Sci. Rep. 5 (2015) 8837.

[4] O. Thiabgoh, T. Eggers, and M. H. Phan, A new contactless magneto-LC resonance technology for real time respiratory motion monitoring, Sens. Actuators A 265 (2017) 120-126.

[5] T. Wang, Y. Zhou, C. Lei, J. Luo, S. Xie, H. Pu, Magnetic impedance biosensor-A review, Biosens. Bioelectron. 90 (2017) 418-435.

[6] D. S. Lam, J. Devkota, N. T. Huong, H. Srikanth, and M.H. Phan, Enhanced high-frequency magnetoresistance responses of melt-extracted Co-rich soft ferromagnetic microwires, J. of Electron. Mater. 45 (2016) 2395-2400.

[7] O. Thiabgoh, T. Eggers, H. Shen, A. Galati, J. Sida, J. F. Sun, H. Srikanth, and M.H. Phan Enhanced high-frequency GMI response of melt-extracted amorphous $Co_{69.25}Fe_{4.25}Si_{13}B_{13.5}$ microwires subject to Joule annealing, J. Sci. Adv. Mater. Devices. 1 (2016) 69-74.

[8] T. Eggers, O. Thiabgoh, S. D. Jiang, H. X. Shen, J. S. Liu, J.F. Sun, H. Srikanth, and M.H. Phan, Tailoring circular magnetic domain structure and high frequency magneto-impedance of melt-extracted $Co_{69.25}Fe_{4.25}Si_{13}B_{13.5}$ microwires through Nb doping, AIP Advances 7 (2017) 056643.

[9] S.D. Jiang, T. Eggers, O. Thiabgoh, D.W. Xing, W. D. Fei, H. X. Shen, J. S. Liu, J.R. Zhang, W.B. Fang, J.F. Sun, H. Srikanth, and M.H. Phan, Relating surface roughness and magnetic domain structure to giant magneto-impedance of Co-rich melt-extracted microwires, Sci. Rep. 7 (2017) 46253.

[10] A. Zhukov, M. Ipatov, V. Zhukova, Advances in giant magnetoimpedance of materials, in K.H.J. Buschow, Handbook of Magnetic Materials 24 (2015) 139-236.

[11] H. X. Peng, F. Qin, M. H. Phan, Ferromagnetic Microwire Composites from Sensors to Microwave Applications, Springer, Switzerland, 2016.

[12] D.-M. Chen, D.-W. Xing, F.-X. Qin, J.-S. Liu, H. Wang, et al., Correlation of magnetic domains, microstructure and GMI effect of Joule-annealed melt-extracted $Co_{68.15}Fe_{4.35}Si_{12.25}B_{13.75}Nb_1Cu_{0.5}$ microwires for double functional sensors, Phys. Status Solidi A 210 (2013) 2515-2520.

[13] B. Shen, A. Inoue, C. Chang, Super high strength and good soft-magnetic properties of (Fe, Co)-B-Si-Nb (Fe, Co)-B-Si-Nb bulk glassy alloys with high glass-forming ability, Appl. Phys. Lett. 85 (2004) 4911.

[14] D. M. Chen, D. W. Xing, F. X. Qin, J. S. Liu, H. X. Shen, et al., Cryogenic Joule annealing induced large magnetic field response of Co-based microwires for giant magneto-impedance sensor applications, J. Appl. Phys. 116 (2014) 053907.

[15] J. Liu, F. Qin, D. Chen, H. Shen, H. Wang, et al., Combined current-modulation annealing induced enhancement of giant magnetoimpedance effect of Co-rich amorphous microwires, J. Appl. Phys. 115 (2014) 17A326.

[16] S. Jiang, D. Xing, J. Liu, H. Shen, D. Chen, et al., Influence of microstructure evolution on GMI properties and magnetic domains of melt-extracted Zr-doped amorphous wires with accumulated DC annealing, J. Alloys Compd. 644 (2015) 180-185.

[17] H. Shen, J. Liu, H. Wang, D. Xing, D. Chen, et al., Optimization of mechanical and giant magneto-impedance (GMI) properties of melt-extracted Co-rich amorphous microwires, Phys. Status Solidi A 211 (2014) 1668-1673.

[18] T. Eggers, D.S. Lam, O. Thiabgoh, J. Marcin, P. Švec, et al., Impact of the transverse magnetocrystalline anisotropy of a Co coating layer on the magnetoimpedance response of FeNi-rich nanocrystalline ribbon, J. Alloy. and Compd., 741 (2018) 1105-1111.

[19] D. de Cos, A. Garcia-Arribas, J. M. Barandiaran, Analysis of magnetoimpedance measurements at high frequency using a microstrip transmission line, Sens. Actuators, A 115 (2004) 368-375.

[20] M. Knobel, P. Allia, C. Gomez-Polo, H. Chiriac and M. Vazquez, Joule heating in amorphous metallic wires, J. Phy. D. Appl. Phy., 28 (1995) 2398-2403.

[21] H. Chiriac, I. Astefanoaei, A model of the DC Joule heating in amorphous wires, Phys. Status Solidi. A 153 (1996) 183-189.

[22] C. Kittel, Introduction to Solid State Physics, eighth ed., Wiley, USA, 2004.

[23] S.-D. Jiang, D.-W. Xing, W.-D. Fei, J.-S. Liu, H.-X. Shen, et al., The disparate impact of two types of GMI effect definition on DC Joule-heating annealed Co-based microwires, RSC Adv. 5 (2015) 103609-103616.

[24] R. C. O'Handley, Modern magnetic materials: principles and applications. John Wiley & Sons, Canada, 2000.

[25] J.M.D. Coey, Magnetism and Magnetic Materials, Cambridge University Press, New York, USA, 2010.

[26] J.G. A. Chizhik, Magnetic Microwires: A Magneto-Optical Study, Pan Stanford, Massachusetts, USA, 2014.

[27] A. Chizhik, A. Stupakiewicz, A. Zhukov, A. Maziewski, J. Gonzalez, Experimental demonstration of basic mechanisms of magnetization reversal in magnetic microwires, Physica B 435 (2014) 125-128.